\documentclass{article}

\usepackage{PRIMEarxiv}

\usepackage[utf8]{inputenc} 
\usepackage[T1]{fontenc}    
\usepackage{hyperref}       
\usepackage{url}            
\usepackage{booktabs}       
\usepackage{amsfonts}       
\usepackage{nicefrac}       
\usepackage{microtype}      
\usepackage{lipsum}
\usepackage{fancyhdr}       
\usepackage{amsmath, algorithm, algpseudocode}
\usepackage{graphicx}       
\usepackage{tikz}
\usepackage{multirow}
\usetikzlibrary{shapes.geometric, arrows, positioning}
\graphicspath{{media/}}     

\title{A Fine-tuning Enhanced RAG System with Quantized Influence Measure as AI Judge
}

\author{
  Keshav Rangan \\
  Columbia University \\
  \texttt{keshavrangan2007@gmail.com} \\
   \And
  Yiqiao Yin \\
  Corresponding Author \\
  University of Chicago, Booth School of Business \\
  Columbia University \\
  \texttt{yy2502@columbia.edu} \\
}

\begin{document}
\maketitle

\begin{abstract}
\begin{quote}
This study presents an innovative enhancement to retrieval-augmented generation (RAG) systems by seamlessly integrating fine-tuned large language models (LLMs) with vector databases. This integration capitalizes on the combined strengths of structured data retrieval and the nuanced comprehension provided by advanced LLMs. Central to our approach are the LoRA and QLoRA methodologies, which stand at the forefront of model refinement through parameter-efficient fine-tuning and memory optimization. A novel feature of our research is the incorporation of user feedback directly into the training process, ensuring the model's continuous adaptation to user expectations and thus, improving its performance and applicability. Additionally, we introduce a Quantized Influence Measure (QIM) as an innovative "AI Judge" mechanism to enhance the precision of result selection, further refining the system's accuracy. Accompanied by an executive diagram and a detailed algorithm for fine-tuning QLoRA, our work provides a comprehensive framework for implementing these advancements within chatbot technologies. This research contributes significant insights into LLM optimization for specific uses and heralds new directions for further development in retrieval-augmented models. Through extensive experimentation and analysis, our findings lay a robust foundation for future advancements in chatbot technology and retrieval systems, marking a significant step forward in the creation of more sophisticated, precise, and user-centric conversational AI systems. We make the \href{https://huggingface.co/datasets/eagle0504/youthless-homeless-shelter-web-scrape-dataset-large}{dataset},  the \href{https://huggingface.co/eagle0504/llama-2-7b-ysa}{model}, and the \href{https://huggingface.co/spaces/eagle0504/YSA-Larkin-Comm}{app} publicly available for the literature.
\end{quote}
\end{abstract}


\section{Introduction}

\noindent The issue of homelessness among families within the United States has escalated into a profoundly serious problem in recent times, as highlighted by Thompson in their 2002 study on the subject. This alarming trend not only underscores the growing socio-economic challenges facing American families but also calls for urgent attention and remedial measures to address this grave concern \cite{thompson2002short}. A significant body of literature has delved into the intricacies and challenges associated with homeless shelters, with a particular focus on addressing the emergency crisis faced by homeless youth. These scholarly works have thoroughly examined the multifaceted issues that contribute to the plight of young individuals without homes, exploring both the immediate and long-term impacts on this vulnerable population segment. Through comprehensive research, these studies aim to shed light on the critical needs and potential interventions required to support homeless youth effectively during their time of crisis \cite{spiegler2022crisis, barber2005homeless, dalton2002adjustment, burt2001helping, thompson2002short, dreyer2018shelter, wallace2018sheltering, hurtubise2009shelters, santos2020elderly}. One of the most urgent issues confronting homeless shelters today is the significant gap in communication channels and the insufficient availability of essential resources \cite{wusinich2019if, hocking2000changing, brown2017waiting, greysen2012understanding}. This problem not only hampers the effective operation of these facilities but also severely restricts the ability of individuals seeking shelter to access the support and services they critically need. The deficiency in clear and open communication pathways within these shelters, coupled with the limited access to necessary resources such as food, healthcare, and counseling services, exacerbates the challenges faced by the homeless population, hindering their journey towards stability and self-sufficiency. \cite{wusinich2019if, hocking2000changing, brown2017waiting, greysen2012understanding, vellozzi2021disparities, barker1990home, haag2011impacting, olufemi2002barriers, haupt2023examining}. 

In this paper, we introduce an innovative approach through the development of a web-based chatbot interface that harnesses the advanced capabilities of Large Language Models (LLMs) and Generative AI (GAI) technologies. Our goal is to significantly enhance the resources available to homeless shelters and the families they serve, thereby substantially improving their channels of communication. While extensive research efforts have been dedicated to refining Large Language Models (LLMs) by fine-tuning them on specialized datasets, there has been a notable lack of focus on applying these advancements to facilitate better communication within the context of homeless shelters. Our project seeks to address this gap by meticulously fine-tuning LLMs using a bespoke dataset, which we have compiled from information sourced from the Youth Spirit Artworks (YSA) Tiny House Empowerment Village website. By doing so, we aim to tailor the LLMs to better meet the specific communication needs of homeless shelters and their residents. Furthermore, in the spirit of fostering wider adoption and facilitating further research, we are committed to making both the dataset and the resulting models publicly accessible. Through this endeavor, we aspire to contribute a meaningful and practical tool that will enhance the communication capabilities of homeless shelters, ultimately benefiting those who rely on these essential services.

\noindent \textbf{Dataset}: \href{https://huggingface.co/datasets/eagle0504/youthless-homeless-shelter-web-scrape-dataset-large}{https://huggingface.co/datasets/eagle0504/youthless-homeless-shelter-web-scrape-dataset}

\noindent \textbf{Model}: \href{https://huggingface.co/eagle0504/llama-2-7b-ysa}{https://huggingface.co/eagle0504/llama-2-7b-ysa}

\noindent \textbf{App}: \href{https://huggingface.co/spaces/eagle0504/YSA-Larkin-Comm}{https://huggingface.co/spaces/eagle0504/YSA-Larkin-Comm}

\section{Related Literature}

\textbf{Fine-tune Large Language Models}: Numerous studies documented within the academic literature have reported considerable success in the process of fine-tuning pre-trained Large Language Models (LLMs) by utilizing Customized Data obtained through extensive internet scraping \cite{he2023large, brown2020language, babaei2023llms4ol, winograd2023loose, yang2023fingpt, ferber2024large, ozdemir2023quick, jamal2023improved}. These research endeavors have meticulously applied adjustments and enhancements to the foundational structures of LLMs \cite{pan2023integrating, kumar2023large, rasnayaka2024empirical}, leveraging the vast array of data available online \cite{babaei2023llms4ol, winograd2023loose, yang2023fingpt, ferber2024large, ozdemir2023quick, jamal2023improved} to tailor these models more closely to specific needs and objectives. This approach has allowed for significant improvements in the performance and applicability of LLMs across a variety of domains, demonstrating the potential of customized datasets to refine and enhance the capabilities of existing language models. These research paved the ground work for the literature and our work took their dedication further to help on the homeless shelters by investigating and scraping the information on the overlooked YSA Homeless Shelter in the Bay Area, California.

To fine-tune pretrained LLMs, there are many data formats such as SQUAD \cite{levy2023guiding, deng2023efficient, ge2023openagi, xue2023repeat} and Guanaco \cite{bekbayev2023poison, dettmers2023qlora} data formats. For text generation tasks, the Guanaco dataset from Open Assistant can be meticulously curated to any text data from internet scraping \cite{dettmers2023qlora}. QLoRA is a finetuning method enabling a 65B model to be finetuned on a 48GB GPU with reduced memory use and preserved performance. The QLoRA techinque uses 4-bit quantized models and Low Rank Adapters, achieving 99.3\% of ChatGPT's performance on the Vicuna benchmark with significant innovations for memory efficiency \cite{dettmers2023qlora}. Many research have demonstrated success in fine-tuning pretrained LLMs using LoRA/QLoRA techniques \cite{li2023loftq, zhang2024quantized, jeon2024l4q, yin2023modulora, zhang2023machine, xu2023qa, guo2023lq, weng2023lmtuner}. Yet, few demonstrated the potential of QLoRA method over small volume of text data scraped from internet. Our investigation show that such fine-tuning strategy can be used on customized low-volume internet data which can support the rising homeless family crisis due to lack of communication and digital support.

It has been demonstrated that extensive pre-trained language models encapsulate factual information within their parameters, leading to unparalleled performance on subsequent NLP tasks when appropriately fine-tuned. Nevertheless, these models face challenges in accurately accessing and manipulating stored knowledge, resulting in their under-performance on tasks that require intensive knowledge compared to specialized task-specific frameworks. Furthermore, the issues of tracing the origins of their decisions and refreshing their repository of world knowledge are yet to be resolved in the field of research \cite{lewis2020retrieval}. A universal fine-tuning methodology for retrieval-augmented generation (RAG) models, which integrate pre-trained parametric and non-parametric memory for the purpose of language generation, has been introduced to the academic community \cite{lewis2020retrieval} and have been receiving noticeable attention by many other researchers \cite{mao2020generation, cai2022recent, liu2020retrieval, gao2023retrieval, jiang2023active}.

\textbf{Retrieval-Augmented Generation}: Another challenge is the inaccuracies amongst the Retrieval-based algorithms. There are intensive research contributed to the literature about the inaccuracies of the Retrieval-Augmented Generation (RAG) algorithm \cite{kim2020retrieval, chen2023benchmarking, li2022survey, goyal2022retrieval, blattmann2022retrieval, siriwardhana2023improving, siriwardhana2023improving, gao2022retrieval, guo2024retrieval}.

\textbf{Influence Measure}: Chernoff, Lo, and Zheng (2009) \cite{chernoffetal2009} presents a general intensive approach, based on a method pioneered by Lo and Zheng (2002) \cite{lozheng2002} for detecting which, out of many potential explanatory variables, have an influence (impact) on a dependent variable $Y$. Related work \cite{lo2021interaction, lo2021novel, lo2021language, di2023detecting} present an interaction-based feature selection methodology incorporating the notion of influence score, I-score, as a major technique to detect the higher-order interactions in complex and large-scale data set. The original measure that assess the predictivity of the a variable set given the response variable is introduced in previous work (for definition of predictivity, see \cite{lo2015significant} and \cite{lo2016framework}).

Suppose there is a response variable $Y$ to be binary (taking values 0 and 1) and all explanatory variables to be discrete. Consider the partition $\mathcal{P}_k$ generated by a subset of $k$ explanatory variables $\{X_{b_1}, ..., X_{b_k}\}$. Assume all variables in this subset to be binary. Then there are $2^k$ partition elements; see the first paragraph of Section 3 in (Chernoff et al., 2009 \cite{chernoffetal2009}). Let $n_1(j)$ be the number of observations with $Y = 1$ in partition element $j$. Let $\bar{n}(j) = n_j \times \pi_1$ be the expected number of $Y = 1$ in element $j$. Under the null hypothesis the subset of explanatory variables has no association with $Y$, where $n_j$ is the total number of observations in element $j$ and $\pi_1$ is the proportion of $Y = 1$ observations in the sample. In Lo and Zheng (2002) \cite{lozheng2002}, the influence score is defined as 
\begin{equation}\label{eq:iscore-raw}
    I(X_{b_1}, ..., X_{b_k}) = \sum_{j \in \mathcal{P}_k} [n_1(j) - \bar{n}_1(j)]^2.
\end{equation}

The statistics I-score is the summation of squared deviations of frequency of $Y$ from what is expected under the null hypothesis. There are two properties associated with the statistics $I$. First, the measure $I$ is non-parametric which requires no need to specify a model for the joint effect of $\{X_{b_1}, ..., X_{b_k}\}$ on $Y$. This measure $I$ is created to describe the discrepancy between the conditional means of $Y$ on $\{X_{b_1}, ..., X_{b_k}\}$ disregard the form of conditional distribution. Secondly, under the null hypothesis that the subset has no influence on $Y$, the expectation of $I$ remains non-increasing when dropping variables from the subset. The second property makes $I$ fundamentally different from the Pearson's $\chi^2$ statistic whose expectation is dependent on the degrees of freedom and hence on the number of variables selected to define the partition. We can rewrite statistics $I$ in its general form when $Y$ is not necessarily discrete
\begin{equation}\label{eq:iscore}
    I = \sum_{j \in \mathcal{P}} n_j^2 (\bar{Y}_j - \bar{Y})^2,
\end{equation}
where $\bar{Y}_j$ is the average of $Y$-observations over the $j^\text{th}$ partition element (local average) and $\bar{Y}$ is the global average. Under the same null hypothesis, it is shown (Chernoff et al., 2009 \cite{chernoffetal2009}) that the normalized $I$, $I/n\sigma^2$ (where $\sigma^2$ is the variance of $Y$), is asymptotically distributed as a weighted sum of independent $\chi^2$ random variables of one degree of freedom each such that the total weight is less than one. It is precisely this property that serves the theoretical foundation for the following algorithm.

\textbf{Contribution of our work}: This paper brings to the literature the following contributions.

\begin{itemize}
    \item The paper introduces a novel system that enhances retrieval-augmented generation (RAG) models by integrating fine-tuned large language models (LLMs) with a traditional vector database, aiming to improve communication within homeless shelters and potentially other communities in need.
    \item It highlights the use of LoRA and QLoRA for parameter-efficient fine-tuning and memory optimization, and introduces a Quantized Influence Measure (QIM) as an "AI Judge" to refine the accuracy and relevance of query result selection.
    \item The study demonstrates the potential of the proposed system to significantly impact not only the AI community by advancing conversational AI technologies but also to serve low-income and underserved populations by providing better access to information and resources.
\end{itemize}

\section{Proposed Method}



\subsection{A Fine-tuning Enhanced Retrieval-Augmented Generation System}

\textbf{Fine-tuning Strategy}: LoRA stands as the most favored and widely utilized Parameter-Efficient Fine-Tuning (PEFT) method, first presented in a 2021 paper. It adopts an adapter-based strategy, integrating additional parameters into the model for training purposes. The innovative aspect lies in the manner these new parameters are incorporated and seamlessly reintegrated into the model, achieving this without expanding the model's total parameter count \cite{aghajanyan2020intrinsic}. LoRA operates by decomposing the matrix responsible for updating weights during training into more manageable, smaller matrices. Imagine a visual where the core matrix, responsible for capturing the adjustments learned through backpropagation, is equivalent in size to the total number of parameters that require modification for fine-tuning the model. This primary matrix can be effectively represented through the use of smaller matrices, identified here as $A$ and $B$, each characterized by a specific rank denoted as $r$. The rank $r$ plays a pivotal role in determining the dimensions of these smaller matrices.

The advantage of this approach lies in the ability to train the model using standard backpropagation techniques, focusing on adjusting the parameters within these compact matrices instead of directly within the entire model framework. Essentially, the learning of the weight updates ($\nabla W$) is facilitated through these diminutive matrices. By subsequently combining these smaller matrices, one can reconstruct the original update matrix. This method significantly reduces the number of parameters involved, thereby lowering computational demands. Moreover, it allows for more efficient storage solutions since only the smaller matrices need to be saved, not the entire model, leading to reduced checkpoint sizes.

QLoRA enhances efficiency by incorporating three innovative techniques aimed at minimizing memory usage without compromising performance quality. These innovations include the 4-bit Normal Float, Double Quantization, and Error Reduction. Let's delve into the specifics of these three crucial advancements \cite{he2023efficientdm}. The 4-bit NormalFloat (NF) introduces an information-theoretically optimal data type utilizing Quantile Quantization techniques. It operates by estimating the $2^k + 1$ quantiles (where $k$ represents the bit count) within a $[0, 1]$ distribution, subsequently normalizing these values to the $[-1, 1]$ interval. Following this, neural network weights are also normalized to the $[-1, 1]$ range and then quantized based on the determined quantiles. Double quantization simplifies the constants used in 4-bit NF quantization, saving an average of 0.5 bits per parameter. QLoRA employs block-wise $k$-bit quantization, segmenting weights into chunks for independent quantization. This approach generates multiple quantization constants, which are further quantized to conserve space, a process feasible due to their minimal count and low computing or storage needs. Quantile quantization groups a broad spectrum of numbers into categories or bins, resulting in various numbers being assigned to the same category. For instance, numbers like 2 and 3 might both be rounded to 3 through quantization, introducing an error of 1 when the weights are later dequantized.



Many research provided the evidence that QLoRA is able to more efficiently assist the fine-tuning workflow in training LLMs \cite{he2023efficientdm, schreiber2023esmbind, zi2023delta, xu2023qa, xia2024chain, dettmers2023qlora}. The proposed fine-tuning algorithm is stated in \ref{algo:fine-tune-qlora} and the parameters fine-tuned are rank ($r$), learning rate ($\alpha$), and dropout rate.

The ``fine-tuning enhanced RAG algorithm" represents a sophisticated approach to augmenting the capabilities of retrieval-augmented generation (RAG) models by incorporating both a traditional vector database and insights from a fine-tuned large language model (LLM) on customized dataset. Initially, the process begins with a PDF document that is processed through PyMuPDF, an efficient library for extracting text from PDFs. This text is then utilized in two distinct pathways: one leads to the creation of a vector database via Chroma, a tool designed to convert text into a searchable vector format, and the other to data processing that prepares the text for fine-tuning an LLM. This dual-pathway approach leverages the strengths of both structured data retrieval and the nuanced understanding of language models.

\begin{figure}[ht]
    \centering
    \resizebox{.47\textwidth}{!}{
        \begin{tikzpicture}[
            node distance=0.5cm and 1.5cm,
            auto,
            block/.style={rectangle, draw, fill=blue!20, text width=6em, text centered, rounded corners, minimum height=4em},
            line/.style={draw, -latex}
        ]
        
        \node[block] (pdf) {PDF};
        \node[block, below=of pdf] (pymupdf) {PyMuPDF};
        \node[block, below=of pymupdf] (chroma) {Chroma};
        \node[block, right=5cm of chroma] (dataprocess) {Data Processing};
        \node[block, below=of chroma] (vectordb) {Vector Database};
        \node[block, below=of dataprocess] (finetune) {Fine-tuned LLM};
        \node[block, below=of vectordb] (query1) {Query: Answer 1};
        \node[block, below=of finetune] (query2) {Query: Answer 2};
        \node[block, right=1.5cm of chroma] (userquestion) {User Question};
        \node[block, right=1.5cm of query1] (combine) {Combine Answers};
        \node[block, below=of combine] (finalmodel) {Foundational LLM};
        \node[block, below=of finalmodel] (finalanswer) {Final Answer};
        
        \path[line] (pdf) -- (pymupdf);
        \path[line] (pymupdf) -- (chroma);
        \path[line] (chroma) -- (vectordb);
        \path[line] (pdf) -| (dataprocess);
        \path[line] (dataprocess) -- (finetune);
        \path[line] (userquestion) -- (vectordb);
        \path[line] (userquestion) -- (finetune);
        \path[line] (vectordb) -- (query1);
        \path[line] (finetune) -- (query2);
        \path[line] (query1) -- (combine);
        \path[line] (query2) -- (combine);
        \path[line] (combine) -- (finalmodel);
        \path[line] (finalmodel) -- (finalanswer);
        
        \end{tikzpicture}
    }
    \caption{\textbf{Executive Diagram for Fine-tuning Enhanced RAG}. The fine-tune enhanced RAG algorithm integrates vector database queries with fine-tuned LLM insights to generate contextually rich and accurate responses.}
    \label{fig:fine-tuning-enhanced-rag}
\end{figure}
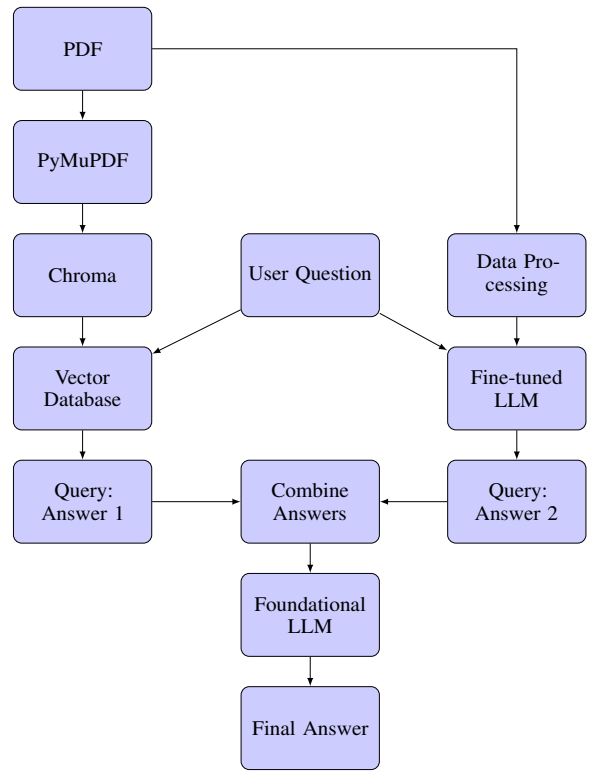

The executive diagram of this approach is presented in Figure \ref{fig:fine-tuning-enhanced-rag}. When a user poses a question, the algorithm engages both pathways to generate comprehensive responses. The question is first used to query the vector database, identifying similar content based on vector similarity, which produces "answer 1". Simultaneously, the same question is fed into the fine-tuned LLM, generating "answer 2" based on the model's learned nuances and understanding from the fine-tuning process. These two answers, each bringing a different perspective and type of insight, are then synthesized. This synthesis involves combining the direct, data-driven response from the vector database with the nuanced, context-aware response from the fine-tuned LLM.

The combined content serves as a set of instructions or relevant content for a foundational LLM, which has not been fine-tuned specifically for this task. This step is crucial as it allows the foundational model to leverage the strengths of both the precise data retrieval from the vector database and the nuanced understanding from the fine-tuned LLM, resulting in a final answer that is both relevant and contextually rich. By integrating these different sources of information and processing, the fine-tune enhanced RAG algorithm significantly improves the ability to generate accurate, detailed, and contextually appropriate answers, pushing the boundaries of what retrieval-augmented models can achieve.

\subsection{Proposed Algorithm for Fine-tune QLoRA}

The fine-tuning strategy for QLoRA outlined in the algorithm is a systematic approach to optimizing the parameters of a model for enhanced performance. Initially, the algorithm begins by setting random values for three critical parameters: the attention dimension ($r$), the LoRA scaling factor ($\alpha$), and the dropout probability for LoRA layers ($\text{dropout}$). These parameters are pivotal in adjusting the model's ability to focus on relevant parts of the input data, scale its learning rate adaptively, and prevent overfitting, respectively. The core of the algorithm operates in a loop that iteratively adjusts these parameters until the validation set error falls below a predefined threshold, signifying satisfactory model performance.

Within this iterative process, the algorithm employs a focused tuning approach by fixing two parameters at a time and varying the third within a predetermined range. This methodical adjustment is done in a sequence: first fixing $r$ and $\alpha$ to optimize $\text{dropout}$, then fixing $r$ and $\text{dropout}$ to find the best $\alpha$, and finally, fixing $\alpha$ and $\text{dropout}$ to adjust $r$. After each parameter adjustment, the algorithm selects the value that yields the lowest validation set error, thereby gradually refining the model's configuration. This cycle repeats until the model achieves an error rate below the set threshold, at which point the fine-tuning concludes. This strategic, step-by-step parameter optimization ensures that each aspect of the model is individually addressed, leading to a comprehensive and efficient fine-tuning process that enhances the model's accuracy and reliability. This is presented in Algorithm \ref{algo:fine-tune-qlora}.

\begin{algorithm}[H]
\caption{Fine-tuning strategy for QLoRA}
\label{algo:fine-tune-qlora}
\begin{algorithmic}[1]
\State Initialize random parameters: $r$, $\alpha$, and $\text{dropout}$, where $r$ is the attention dimension, $\alpha$ is the LoRA scaling factor, and $\text{dropout}$ is the dropout probability for LoRA layers.
\While{validation set error $>$ threshold}
  \State Fix $r$ and $\alpha$. Vary $\text{dropout}$ within a predefined range and select the value that minimizes the validation set error. Set $\text{dropout}$ to this optimal value.
  \State Fix $r$ and $\text{dropout}$. Vary $\alpha$ within a predefined range and select the value that minimizes the validation set error. Set $\alpha$ to this optimal value.
  \State Fix $\alpha$ and $\text{dropout}$. Vary $r$ within a predefined range and select the value that minimizes the validation set error. Set $r$ to this optimal value.
\EndWhile
\State \textbf{end training}
\end{algorithmic}
\end{algorithm}

\subsection{Proposed Quantized Influence Measure as AI Judge}

\begin{figure}[ht]
    \centering
    \begin{tabular}{cc}
        \includegraphics[width=.5\textwidth]{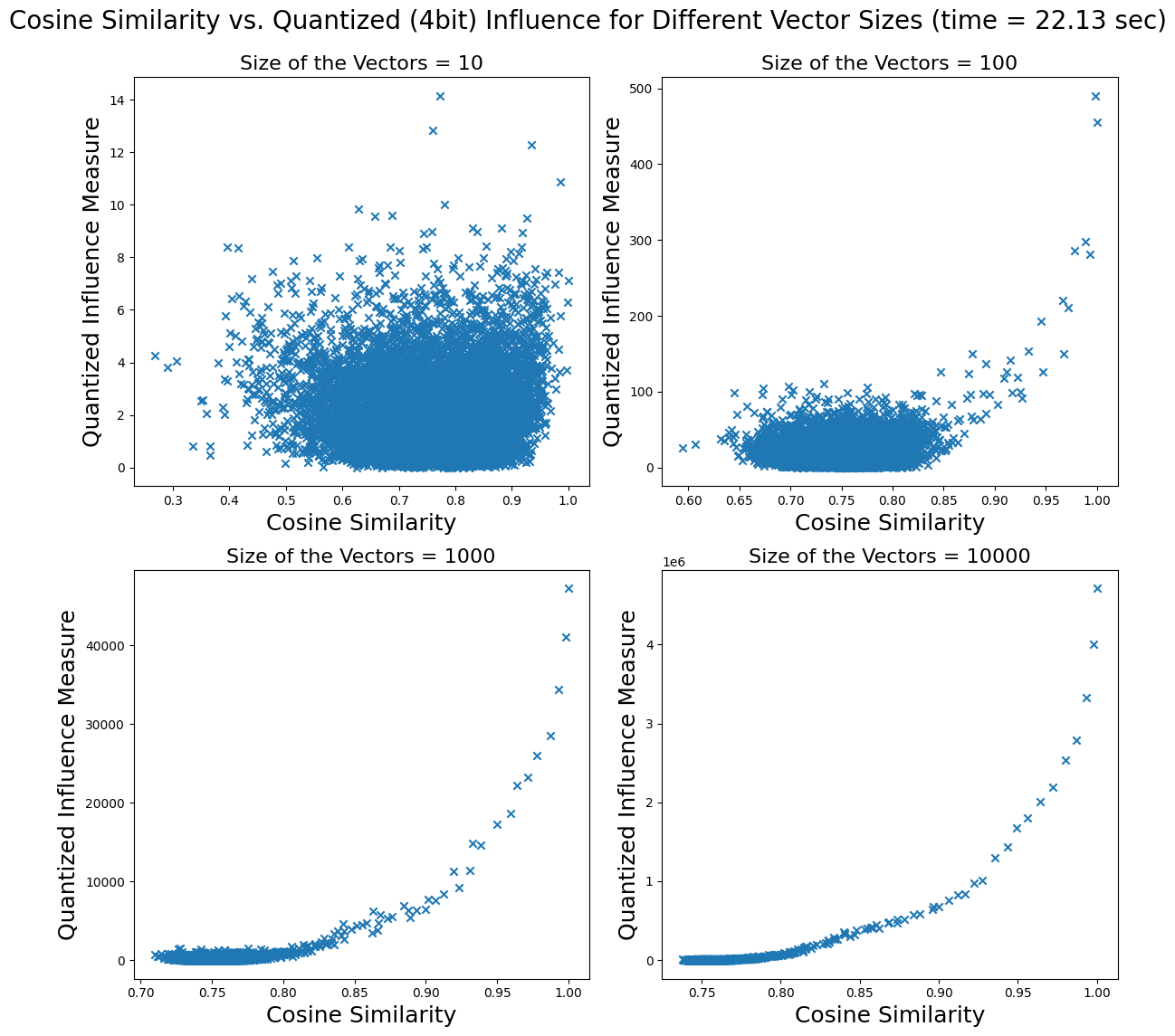} &
        \includegraphics[width=.5\textwidth]{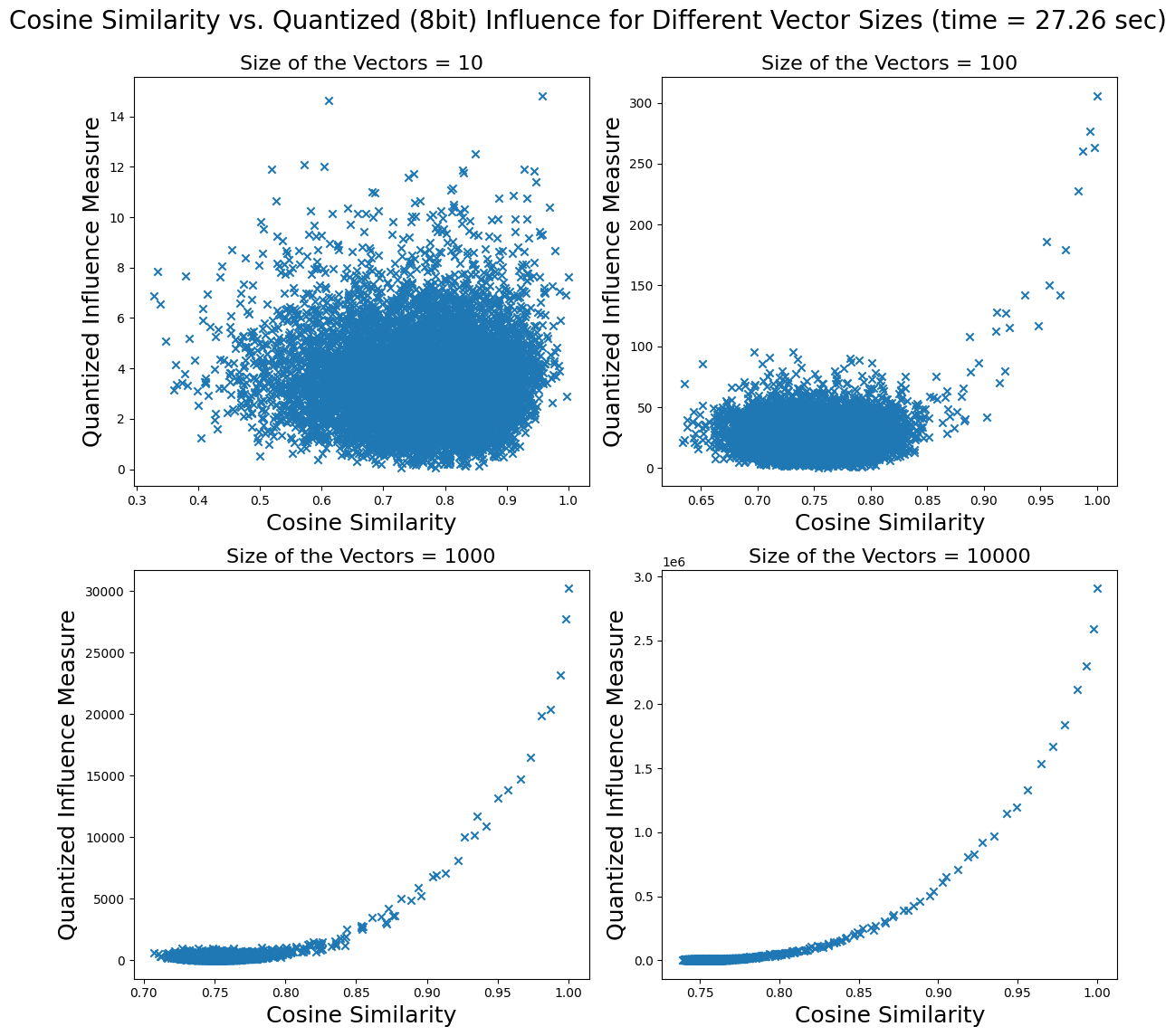} \\ \includegraphics[width=.5\textwidth]{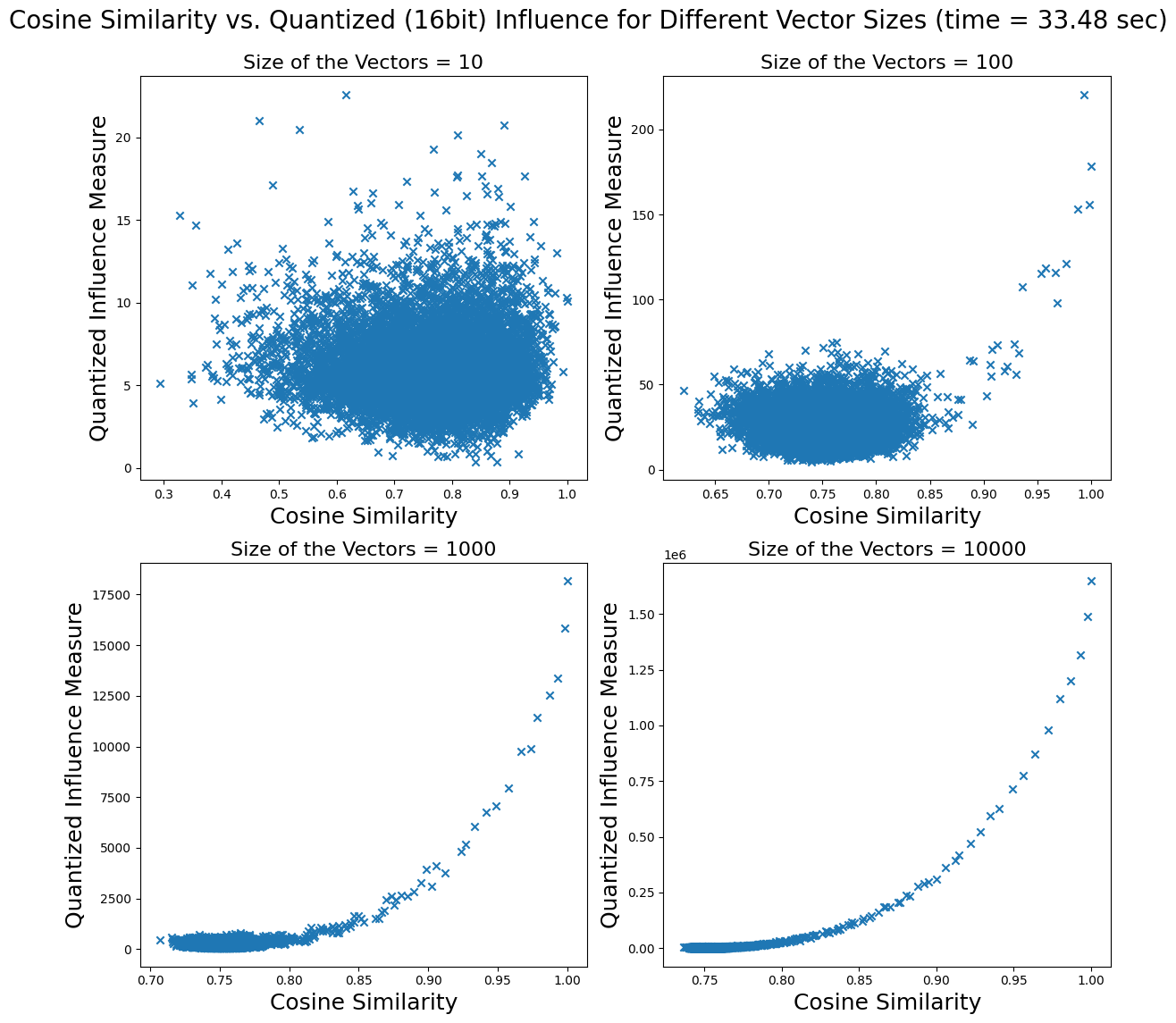} & \includegraphics[width=.5\textwidth]{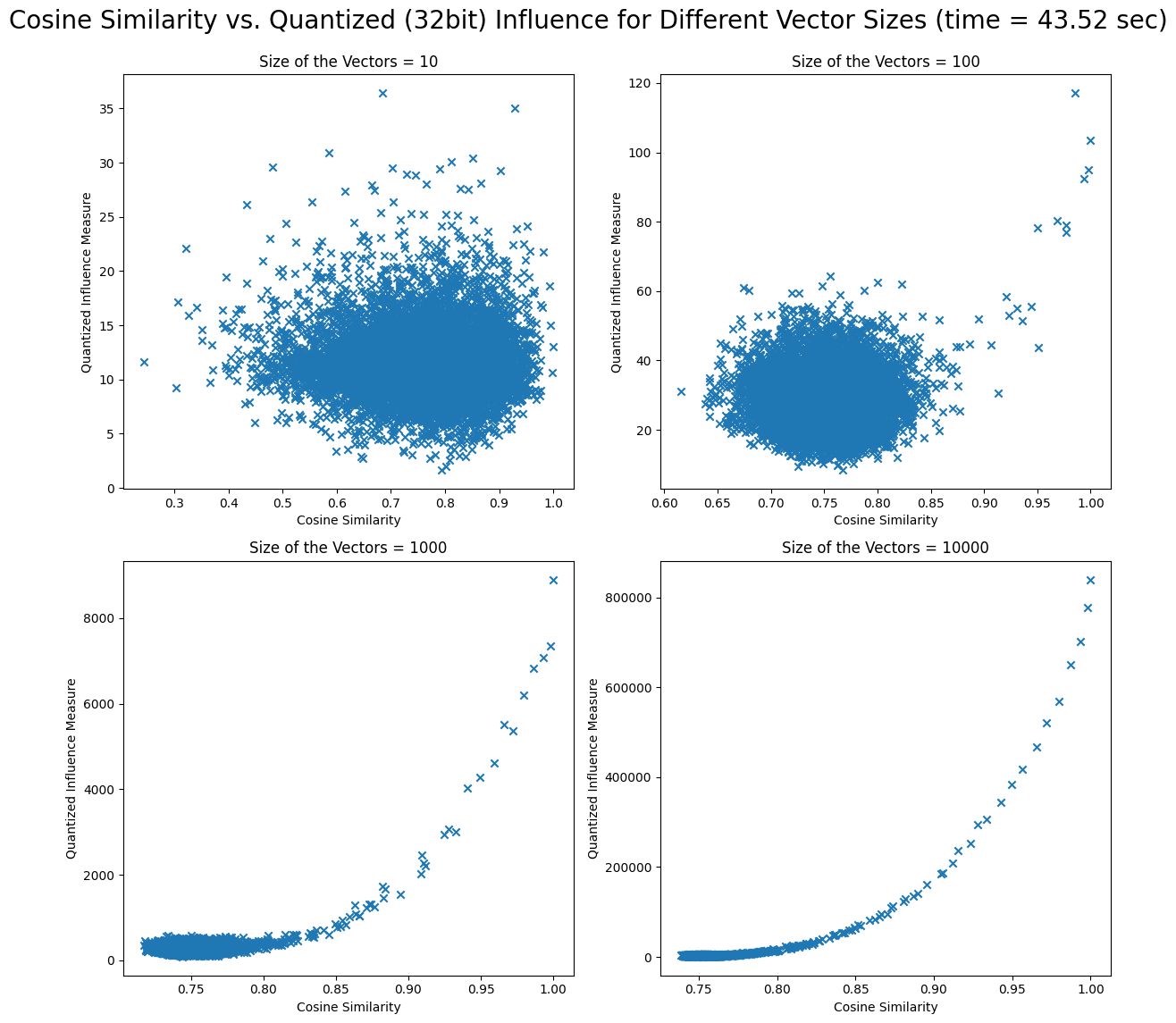} \\
    \end{tabular}
    \caption{\textbf{Comparison of the behavior between cosine and QIM}. Graphical analysis of how vector size affects the relationship between cosine similarity and quantized influence measure. For vectors of size 10, we observe that the signals are random. However, in practice the embedding layers produce vector representation of size 1000 or above. The simulation shows that for vectors of size 1000 the value of quantized influence measure increases exponentially. For the extremely high similarity content, it is much easier to use quantized influence measure to filter and select the relevant content/reference in the RAG algorithm. The quantized concept is a tuning parameter and the experiment shows $q$-bit can be changed from 4 to 32, i.e. delivering better results but with longer time consumption. To select the $q$ parameter, it is worth noting that the higher $q$ values lead to more densely generated partitions, but the calculation of the Quantized Influence Measure would also take longer time. }
    \label{fig:toy-example}
\end{figure}

\textbf{Quantized Influence Measure}: Suppose the goal is to measure the similarity of two arrays, i.e. a query $X$ and a reference $Y$. The Quantized Influence Measure (QIM) computes a score based on the difference in local averages from the global average of $Y$, weighted by the square of the local average multiplied by count of elements in each partition and normalized by the standard deviation of $Y$. The formula for the quantized influence can be expressed as follows:

\begin{equation}\label{eq:quantized-influence-measure}
    \text{Quantized Influence} = \frac{\sum_{i=1}^{q} \left( \overline{y}_{\text{local}, i} - \overline{y}_{\text{global}} \right)^2 \cdot N_i^2}{q \cdot \sigma_{Y}}
\end{equation}

where $\overline{y}_{\text{local}, i}$ is the local average of $Y$ for the $i^{th}$ unique value in $X$, $\overline{y}_{\text{global}}$ is the global average of $Y$, $N_i$ is the count of elements in $Y$ that correspond to the $i^{th}$ unique value in $X$, $\sigma_{Y}$ is the standard deviation of $Y$, $q$ is the total number of unique values in $X$. The equation \ref{eq:quantized-influence-measure} provides a general form where $i$ is the running index of the array $X$ and assuming the array is from real numbers ($\mathbb{R}$) there can be possibly $n$ values. The quantized concept is a tuning parameter and the experiment (see Figure \ref{fig:toy-example}) shows $q$-bit can be changed from 4 to 32, i.e. delivering better results but with longer time consumption. It is recommended to use simulation to guide the selection of this parameter under the committed computing resources at hands.

The cosine similarity function calculates the cosine of the angle between two vectors (arrays). The formula for cosine similarity is:

\begin{equation}\label{eq:cosine-similarity}
    \text{Cosine Similarity} = \frac{\vec{a} \cdot \vec{b}}{\| \vec{a} \| \cdot \| \vec{b} \|}
\end{equation}

where $\vec{a}$ and $\vec{b}$ are the vectors corresponding to `arr1` and `arr2`, respectively, and $\| \vec{a} \|$ and $\| \vec{b} \|$ are the Euclidean norms (magnitudes) of vectors $\vec{a}$ and $\vec{b}$, respectively. In the case of measuring similarity between a query and a reference, we can consider the two arrays to be $X$ and $Y$.

\textbf{Effect of the power term in the quantized influence measure}. We argue that the power term in the quantized influence measure (defined in equation \ref{eq:quantized-influence-measure}) can provide large numerical value to truly single out the ``extremely" similar and relevant content for user than the cosine similarity. We discuss the effect of the term $N_i^2$ in the quantized influence measure formula on making its measure exponentially higher than the cosine similarity, especially as the numerical measure gets higher, we will focus on the impact of this term. The $N_i^2$ term in the quantized influence measure formula significantly increases the influence of partitions with more elements. As the sample size (or the number of elements corresponding to a unique value in $X$) increases, the $N_i^2$ term grows quadratically, making the overall quantized influence measure potentially much larger, especially for data sets where some values in $X$ correspond to many more elements in $Y$ than others.

To illustrate the exponential increase and compare the two measures, let us consider the scenario where the sample size goes to infinity. We simplify the scenario to focus on the effect of the $N_i^2$ term. Suppose that the local averages and global averages remain constant, and we ignore the normalization by standard deviation for simplicity. For quantized influence measure, as $N_i$ increases, the term $N_i^2$ will dominate the measure, causing it to increase quadratically. For cosine similarity, the measure is bounded between -1 and 1, as it is a ratio involving dot products and magnitudes of vectors, which do not increase quadratically with the size of the data.

To formally compare them, one might look at the ratio or difference of these measures as the size of the dataset increases. However, given that cosine similarity is bounded and quantized influence measure increases with $N_i^2$, any direct comparison would show that the influence measure grows significantly faster and larger than the cosine similarity as the dataset size increases, underlining the quadratic impact of $N_i^2$. This demonstrates conceptually why the quantized influence measure measure could exponentially exceed cosine similarity as numerical measures get higher, particularly due to the quadratic growth contributed by the $N_i^2$ term. A formal proof would involve defining specific behaviors for the averages and distributions of `arr1` and `arr2`, which goes beyond this conceptual explanation. In practice, these arrays can be considered as a query $X$ and a reference $Y$ and cosine similarity is a common practice in the literature whereas in our work we propose to use quantized influence measure (QIM).

To present a formal proof comparing the exponential increase of the quantized influence measure measure relative to the cosine similarity measure, let's simplify and focus on key aspects of each formula, especially emphasizing the impact of the $N_i^2$ term in quantized influence measure.

We list the following assumptions:
\begin{itemize}
    \item The cosine similarity is bounded between $[-1, 1]$ due to its definition.
    \item The local average difference squared $\left( \overline{y}_{\text{local}, i} - \overline{y}_{\text{global}} \right)^2$ in the quantized influence measure formula can be considered constant $C$ for simplification.
    \item $N_i$ represents the size of partitions, and we let it approach infinity to analyze the impact.
\end{itemize}

To understand whether a reference from the RAG system provides similar content to the user prompt or not, it is important to filter the extremely ``relevant" content. The common way is to use the cosine similarity (defined in equation \ref{eq:cosine-similarity}). However, we show that as $N_i$ (the size of partitions in $Y$ for each unique value in $X$) approaches infinity, the quantized influence measure measure increases at a rate that is significantly higher than any possible value of cosine similarity. Given the simplified quantized influence measure formula without normalization by standard deviation for illustration:

\begin{equation}\label{eq:quantized-influence-measure-simplified}
    \text{Quantized Influence} = \frac{\sum_{i=1}^{n} C \cdot N_i^2}{q}
\end{equation}

Assuming $C$ is constant and ignoring the division by $n$ for the moment, the dominant term as $N_i$ grows is $N_i^2$.

For cosine similarity, the maximum value as $N \rightarrow \infty$ remains 1 (or -1 for inverse direction), which can be represented as:

\begin{equation}\label{eq:cosine-similarity-limit}
    \lim_{N \rightarrow \infty} \text{Cosine Similarity} = 1
\end{equation}

For quantized influence measure, as $N_i$ increases:

\begin{equation}\label{eq:quantized-influence-measure-limit}
    \lim_{N_i \rightarrow \infty} \text{Quantized Influence} = \lim_{N_i \rightarrow \infty} C \cdot N_i^2
\end{equation}

Since $C$ is a positive constant and $N_i^2$ increases quadratically:

\begin{equation}\label{eq:quantized-influence-measure-limit-final}
    \lim_{N_i \rightarrow \infty} C \cdot N_i^2 = \infty
\end{equation}

The quantized influence measure measure grows without bound as the size of the partitions $N_i$ increases, particularly because of the $N_i^2$ term, which ensures that this growth is quadratic. In contrast, cosine similarity is inherently limited to a maximum value of 1, regardless of the size of the input vectors.

This demonstrates that as the partition sizes $N_i$ increase, the difference between the quantized influence measure measure and the cosine similarity measure not only grows but does so in a manner that can be considered exponential due to the quadratic factor of $N_i^2$. Hence, we've shown that the quantized influence measure measure will be strictly larger than the cosine similarity measure as $N_i$ (and thereby the sample size) goes to infinity, highlighting the significant impact of the $N_i^2$ term in the former measure.

\textbf{A Toy Example}: The experiment investigates the dynamics between cosine similarity and quantized influence metrics across increasing vector dimensions, specifically for sizes $n = 10, 100, 1000, \text{and } 10000$. For each vector size, pairs of vectors are generated, where one vector, $\mathbf{a}$, serves as a baseline, and the other vector, $\mathbf{b}$, is its perturbed counterpart. The perturbation involves adding a scaled random vector to $\mathbf{a}$, mathematically represented as $\mathbf{b} = \mathbf{a} + k \cdot \text{rand}(n)$, with $k$ varying to introduce different levels of deviation and $\text{rand}(n)$ producing a vector of $n$ uniformly distributed random numbers, i.e. $\text{rand}(n) \sim \mathcal{U}(0,1)$

Cosine similarity and quantized influence between $\mathbf{a}$ and $\mathbf{b}$ are calculated for each perturbation level, plotted against each other to analyze how these relationships evolve with vector size. Please see Figure \ref{fig:toy-example}. This methodology enables a detailed examination of the interplay between similarity and influence in vector spaces, particularly how dimensionality influences the sensitivity and behavior of these metrics under varying degrees of vector modification. Through this analytical framework, the study aims to elucidate the underlying patterns and principles governing vector relationships in high-dimensional data analysis, contributing valuable insights into the nature of similarity and influence within complex vector spaces.

As the perturbation factor increased, the study observed the impact on both metrics, capturing their values across a spectrum of perturbation levels (see Figure \ref{fig:toy-example}). This approach allowed for a nuanced understanding of how changes in vector composition affect their perceived similarity and influence, providing insights into the robustness and sensitivity of these metrics to alterations in vector content and size. The findings are visualized in a series of scatter plots, each corresponding to a different vector size, illustrating the complex interplay between cosine similarity and quantized influence as vector dimensions escalate. This analysis sheds light on the underlying dynamics of vector similarity and influence in high-dimensional spaces, contributing to the broader discourse on vector analysis and its applications in data science and machine learning.

\subsection{Proposed System in Production}

\begin{figure}[ht]
    \centering
    \includegraphics[width=1\textwidth]{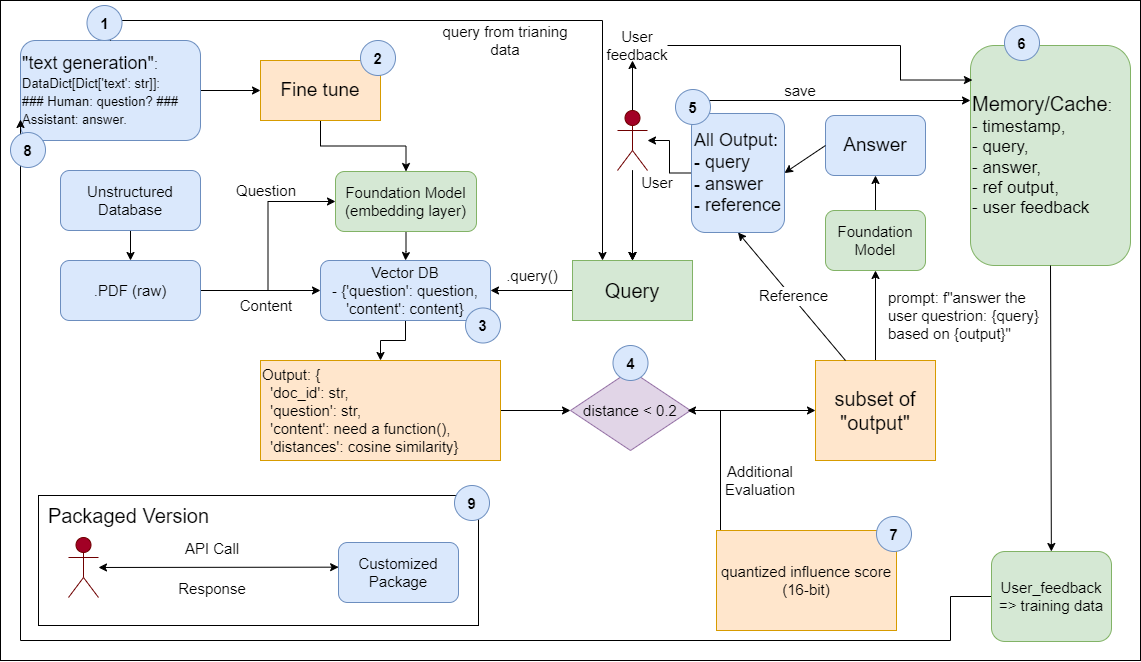}
    \caption{\textbf{Proposed System Architect}. This executive diagram explains the system architecture to implement the proposed method in a chatbot. (1) The training data is created using the ``text-generation" style. This is a dictionary with ``Human" and ``Assistant" referring to question-answer pairs. This gives us the training data for fine-tuning models. (2) We fine tune large language models based on proposed methods discussed in the previous sections (using Algorithm \ref{algo:fine-tune-qlora}). (3) A vector database is created and this collection stores data in the vector form (we use ``chroma" library). (4) The query search the vector database against the user's question/prompt and return selections with distance score. This allows us to filter against a certain threshold, i.e. 0.2. (5) We take the question/prompt, the answer and the reference and display that on the screen for the user. (6) We can ask for user feedback and save the cache to a directory for next-stage training purpose, because we can train another model to learn the user preference. (7) We use the proposed quantized influence measure as an additional ``AI Judge" to help us rank the results in the fourth step. (8) We use the feedback provided from the user to enhance the training data in the first step. (9) In the end, the last step proposes to package the code into one API and have a cleaner version in one software package for technical user to have programmatic access.}
    \label{fig:executive-diagram-for-chatbot}
\end{figure}

This comprehensive executive diagram (see Figure \ref{fig:executive-diagram-for-chatbot}) elucidates the intricate system architecture designed to implement the proposed method within a chatbot framework, aiming to revolutionize how information is processed and delivered in real-time interactions. Initially, the foundation of this innovative approach begins with the creation of training data, meticulously crafted in a "text-generation" style. This involves compiling a dictionary of question-answer pairs labeled as "Human" and "Assistant," respectively. Such a structured dataset is pivotal for the subsequent fine-tuning of large language models (LLMs), ensuring they are aptly prepared to understand and respond to user queries with high accuracy and relevance.

Following the data preparation, the process advances to fine-tune LLMs, employing the sophisticated techniques outlined in previous discussions and specifically referenced through Algorithm 1. This fine-tuning phase is crucial for adapting the models to the nuances and specificities of the targeted application, thereby enhancing their performance and utility in real-world scenarios. Concurrently, a vector database is constructed using the "chroma" library, which serves as a repository for storing data in vector form. This database is instrumental in facilitating efficient and precise query searches against user questions or prompts, employing distance scores to gauge relevance and filter responses based on a predefined threshold, such as 0.2.

Within the meticulously designed system architecture lies an advanced dynamic interaction component. This integral feature meticulously curates and presents the query outcomes, aligning them with the pertinent questions or prompts and associated references directly to the user interface. This immediate and interactive feedback mechanism serves a dual purpose. Primarily, it enlightens the user with real-time information and insights, bridging the gap between query initiation and result delivery. Secondly, it actively solicits user engagement through a structured feedback system. This interaction is not merely superficial; the feedback collected is of paramount importance, as it is methodically cataloged and leveraged in subsequent training cycles. This strategic incorporation of user feedback facilitates the evolution and refinement of the model, ensuring its adaptation and optimization in alignment with genuine user preferences and interaction patterns.

Further enhancing the complexity and effectiveness of this system is the deployment of an "AI Judge." This innovative component utilizes a sophisticated quantized influence measure, a proposal set forth to augment the precision in the ranking process of query outcomes. By integrating this measure, the system introduces a nuanced layer of analysis, significantly elevating the sophistication and accuracy of the result selection mechanism. The feedback garnered from users, following their interaction with the query outcomes, is meticulously integrated back into the foundational training dataset. This process not only enriches the existing dataset but also contributes to the nuanced fine-tuning of the chatbot’s response mechanisms. Through this cyclical enhancement, the architecture achieves a harmonious balance between user-centric customization and algorithmic precision, ensuring the chatbot evolves continually to meet and exceed user expectations.

Finally, the culmination of this extensive process is the packaging of the entire system into a singular API, offering a streamlined and user-friendly software package. This allows technical users to access the enhanced Retrieval-Augmented Algorithm system programmatically, facilitating ease of integration and application in a wide array of settings. Through this meticulously designed system architecture, the proposed method stands to significantly advance the capabilities of chatbots, offering more accurate, relevant, and user-tailored interactions.

\section{Experiment and Discussion}

\subsection{YSA Document Data}

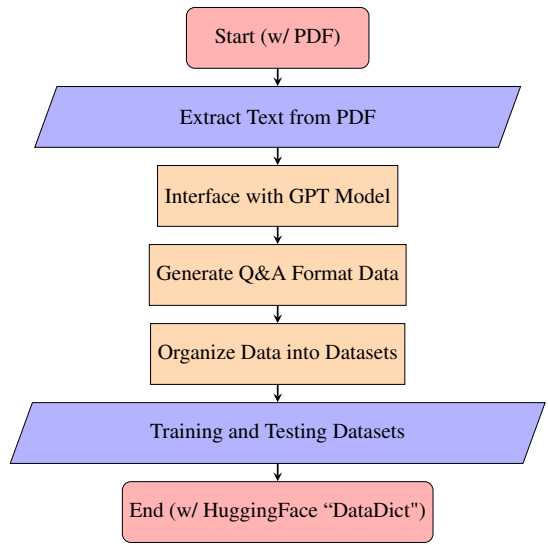
\begin{figure}[ht]
    \centering
    \resizebox{.44\textwidth}{!}{
        \begin{tikzpicture}[node distance=1.3cm]
        
        \tikzstyle{startstop} = [rectangle, rounded corners, minimum width=3cm, minimum height=1cm,text centered, draw=black, fill=red!30]
        \tikzstyle{io} = [trapezium, trapezium left angle=70, trapezium right angle=110, minimum width=3cm, minimum height=1cm, text centered, draw=black, fill=blue!30]
        \tikzstyle{process} = [rectangle, minimum width=3cm, minimum height=1cm, text centered, draw=black, fill=orange!30]
        \tikzstyle{decision} = [diamond, minimum width=3cm, minimum height=1cm, text centered, draw=black, fill=green!30]
        \tikzstyle{arrow} = [thick,->,>=stealth]
        
        \node (start) [startstop] {Start (w/ PDF)};
        \node (in1) [io, below of=start] {Extract Text from PDF};
        \node (pro1) [process, below of=in1] {Interface with GPT Model};
        \node (pro2) [process, below of=pro1] {Generate Q\&A Format Data};
        \node (pro3) [process, below of=pro2] {Organize Data into Datasets};
        \node (out1) [io, below of=pro3] {Training and Testing Datasets};
        \node (stop) [startstop, below of=out1] {End (w/ HuggingFace ``DataDict")};
        
        \draw [arrow] (start) -- (in1);
        \draw [arrow] (in1) -- (pro1);
        \draw [arrow] (pro1) -- (pro2);
        \draw [arrow] (pro2) -- (pro3);
        \draw [arrow] (pro3) -- (out1);
        \draw [arrow] (out1) -- (stop);
        
        \end{tikzpicture}
    }
    \caption{\textbf{Executive Diagram for Data Processing}. This diagram illustrates the process flow from extracting text from PDF documents to generating and organizing conversational Q\&A data for fine-tuning Large Language Models (LLMs).}
    \label{fig:exec-diagram-data-processing}
\end{figure}

The methodology for transforming PDF documents into a format conducive to fine-tuning Large Language Models (LLMs) begins with the extraction of text content using the PyMuPDF library, specifically chosen for its robust handling of PDF files. The executive diagram for data processing is presented in Figure \ref{fig:exec-diagram-data-processing}. The process involves the systematic opening of the PDF document, iterating through each page, and collecting the text content into a comprehensive list. This list, consisting of strings where each string represents the text from a single page, serves as the foundational dataset for subsequent processing steps. This initial stage is crucial for ensuring that all textual information within the document is accurately captured and made available for further manipulation.

Subsequent to text extraction, the methodology advances to interfacing with OpenAI's language models to generate conversational data from the extracted text. This is achieved through a function designed to simulate a dialogue with the GPT model, framing the AI as a helpful assistant. By embedding the text within a conversational context, this function solicits contextually relevant responses from the model, effectively transforming the static text into a dynamic question-answer format. This step is instrumental in creating a dataset that mirrors the interactive nature of conversational AI, enhancing the model's ability to engage in and respond to human-like dialogue. To execute this step, there is an API call to ``ChatGPT'' with customized prompt that encapsulates text data within a specific prompt structure, and then utilizes the LLM to generate a Q\&A pair. The prompt instructs the model to produce content where each line contains a question (marked by ``\#\#\# Human:") followed by an answer (marked by ``\#\#\# Assistant:"), based on the provided content. This structured approach is crucial for generating training data that is directly applicable to enhancing the LLM's conversational capabilities.

The process of enhancing the dataset involves an intricate function specifically designed to organize the generated content into a meticulously standardized question-and-answer (Q\&A) format. This sophisticated function meticulously crafts prompts that adeptly guide the artificial intelligence system to generate outputs where each entry is composed of a question immediately followed by its corresponding answer, meticulously extracted from the underlying text content. This methodical structuring is paramount for curating a dataset that seamlessly aligns with the intricate training prerequisites of Large Language Models (LLMs), thereby facilitating the cultivation of sophisticated conversational abilities within the AI model.

By placing a strong emphasis on adopting a Q\&A format, this approach plays a crucial role in significantly enhancing the model's comprehension of varying contexts, thereby substantially improving its proficiency in generating responses that are not only coherent but also highly relevant to the posed questions. The Q\&A format serves as a vital framework that simulates real-world conversational dynamics, enabling the model to better understand the intricacies of human dialogue. This format ensures that the AI is trained on a dataset that mirrors natural language use, thus equipping the model with the ability to handle a wide range of conversational scenarios. Through this refined training methodology, the model is adeptly prepared to engage in more nuanced and meaningful interactions, demonstrating a deeper understanding of both the questions posed and the appropriate contextually relevant responses.

The final stage of the methodology involves the compilation and organization of the generated Q\&A pairs into distinct datasets for training and testing purposes. By systematically iterating through the text content and applying the specialized function to generate multiple Q\&A pairs from each text segment, a rich dataset is created. This dataset is then divided into training and testing subsets, encapsulated within a ``DatasetDict" object for efficient management. This structured approach to dataset creation is essential for preparing the data in a manner that is optimally suited for fine-tuning LLMs. By providing a diverse and comprehensive set of conversational interactions, the dataset facilitates the enhancement of the LLMs' performance, ensuring they are better equipped to handle a wide array of conversational tasks in real-world applications.

Table \ref{tab:doc-id} presents an organized summary of the overall performance of documents that have been collected and categorized for a specific analysis or project. It details the structure of the dataset by listing document IDs alongside their corresponding names, ranging from "About YSA" to "Application Process," and finally, "Overview." This table efficiently indexes seven distinct segments of the raw data scraped from the internet, each designated with a unique ID from 1 to 7. These documents are systematically arranged to facilitate easy reference and cross-referencing, as indicated in the table's caption. The caption also notes that these partitions are crucial for understanding the broader dataset, implying that each segment plays a specific role in the overall analysis or project framework. The use of such a table underscores the importance of methodical data organization in enhancing the accessibility and interpretability of collected information.

\begin{table}[ht]
    \centering
    \resizebox{0.25\textwidth}{!}{
        \begin{tabular}{cl}
            \hline
            ID & Name \\
            \hline
            1 & About YSA \\
            2 & Board of Directors \\
            3 & Definition of Homeless \\
            4 & Our Team \\
            5 & Programs \\
            6 & Application Process \\
            7 & Overview \\
            \hline
        \end{tabular}
    }
    \caption{\textbf{Document ID}. We partition the raw data scraped from the internet into different partitions. These documents are indexed from 1 to 7, which can also be referenced in Table \ref{tab:overall-results}.}
    \label{tab:doc-id}
\end{table}

\subsection{Results and Discussion}

Table \ref{tab:overall-results} meticulously outlines the outcomes of a series of experiments focused on evaluating a spectrum of models for chatbot creation, featuring Davinci002 (Dav.), fine-tuned on a proprietary dataset; Llama2 with 7 billion parameters (Llam.), similarly fine-tuned; Langchain + SerpAPI (L+S), a novel approach leveraging internet search for information retrieval; a Foundation Model (FM); and three distinct configurations of the Retrieval-Augmented Generation (RAG) system (1E, 3E, and L), each enhanced in unique ways. The embedding model (E) is a choice of selection. One embedding and three embeddings are represented using 1E and 3E, respective. The letter ``L" in RAG (L) refers to RAG system enhanced by Llama2 model. The letter ``L+QIM" in RAG (L+QIM) means the RAG system enhanced with fine-tuned Llama2 model and AI Judge implemented that uses the quantized influence measure as an additional security to ensure the similarity of prompt and reference content. The formula of quantized influence measure is discussed in equation \ref{eq:quantized-influence-measure}. Across seven trials, the table records individual performance scores for these models, culminating in average (Ave.) scores and standard deviations (SD) to convey overall effectiveness and reliability. Notably, the RAG (L) model emerges as a top performer with an impressive average score of 0.934, outshining the Davinci002 and Llama2 models, which post averages of 0.737 and 0.871, respectively. This table also highlights the variability in performance, as indicated by the standard deviation, with RAG models demonstrating remarkable consistency, particularly RAG (L) with an SD of 0.016, suggesting a stable performance across different settings.

A deeper dive into the numerical data reveals the RAG models' dominance over traditional and innovative approaches alike. For example, in specific trials, the RAG (L) model achieved scores as high as 0.960, surpassing other models by a significant margin. The closest competitors, RAG (1E) and RAG (3E), also exhibit strong performances with average scores of 0.925 and 0.921, respectively, indicating the effectiveness of retrieval-augmented strategies. In comparison, the Foundation Model (FM), despite a robust average of 0.848, and the Langchain + SerpAPI (L+S) approach, with a lower average of 0.495, illustrate the challenging nature of achieving high performance in chatbot creation. This contrast is particularly evident when considering the highest scores of traditional models like Davinci002 and Llama2, which barely reach the lower threshold of RAG (L)'s performance range. The system with Llama2 and AI Judge shows the best performance. 

The standout RAG (L) model, an amalgamation of the RAG framework and the Llama2 model, fine-tuned on proprietary data, not only showcases the highest average score but also the most consistent performance across trials, as evidenced by its minimal standard deviation. This precision, coupled with its peak score of 0.960, underscores the synergistic power of combining advanced generative capabilities with targeted, retrieval-augmented mechanisms. The substantial lead of RAG (L) over foundational approaches, including the innovative yet less consistent Langchain + SerpAPI method, highlights the critical importance of integrating contextual retrieval into generative models. This integration significantly enhances the chatbot's responsiveness and accuracy, setting a new benchmark for chatbot technology as demonstrated in these comprehensive experiments.

\begin{table}[ht]
    \centering
    \resizebox{.8\textwidth}{!}{
        \begin{tabular}{ccccccccc}
        \hline
        \hline
            ID	&	Dav.	&	Llam.	&	L+S	&	FM	&	RAG (1E)	&	RAG (3E)	&	RAG (L)  & RAG (L+QIM)	\\
            \hline
        1	&	0.744	&	0.950	&	0.470	&	0.853	&	0.909	&	0.913	&	0.920   &	0.940	\\
        2	&	0.757	&	0.860	&	0.520	&	0.674	&	0.911	&	0.929	&	0.930   &	0.950	\\
        3	&	0.779	&	0.880	&	0.496	&	0.857	&	0.947	&	0.945	&	0.950   &	0.970	\\
        4	&	0.784	&	0.870	&	0.540	&	0.883	&	0.911	&	0.884	&	0.920   &	0.930	\\
        5	&	0.752	&	0.860	&	0.480	&	0.899	&	0.909	&	0.932	&	0.940   &	0.950	\\
        6	&	0.617	&	0.830	&	0.450	&	0.910	&	0.979	&	0.937	&	0.960   &	0.970	\\
        7	&	0.724	&	0.850	&	0.510	&	0.857	&	0.908	&	0.906	&	0.920   &	0.940	\\
        \hline
        Ave.     &	0.737	&	0.871	&	0.495	&	0.848	&	0.925	&	0.921	&	0.934    &	0.950    	\\
        SD          &	0.056	&	0.038	&	0.031	&	0.080	&	0.028	&	0.021	&	0.016 & 0.014	\\
        \hline
        \end{tabular}
    }
    \caption{\textbf{Executive Summary of Results}. The table presents overall results of all candidates that may be used to create chatbot. The document IDs are defined in Table \ref{tab:doc-id}. The performance is measured using cosine similarity. The average (Ave.) and the standard deviation (SD) are also displayed in the table. }
    \label{tab:overall-results}
\end{table}

\subsection{Discussion of Fine-tuning Llama2 on Proprietary Dataset}

The table elucidates the meticulous process of fine-tuning parameters for an algorithm designed to optimize chatbot creation, detailed across three distinct panels labeled A, B, and C. Each panel concentrates on adjusting a single parameter—dropout in Panel A, alpha in Panel B, and the attention dimension (r) in Panel C—while keeping the others constant. This sequential tuning method aims to methodically reduce the loss, thereby enhancing the model's performance. For instance, Panel A experiments with dropout values of 0.001, 0.01, and 0.1, observing minimal variation in loss, indicating a relatively stable performance across these settings. Panel B shifts focus to the alpha parameter, demonstrating a more pronounced effect on loss reduction as alpha increases from 8 to 64, with the lowest loss noted at 0.1122. Panel C, adjusting the attention dimension, corroborates the minimal impact on loss, with values tightly clustered around 0.1122, showcasing an effective fine-tuning strategy that culminates in a significant loss reduction from initial experiments.

Delving deeper into the numerical details reveals the fine-tuning's efficacy. In Panel A, the dropout parameter is finely adjusted, yet the loss remains fairly consistent, suggesting that changes in dropout have a marginal effect on the model's loss within the tested range. Transitioning to Panel B, where alpha is varied from 8 to 64, a clear trend emerges: as alpha increases, the loss significantly decreases, culminating in a remarkable 69\% reduction in loss from the highest (0.5904) to the lowest recorded value (0.1122). This suggests that increasing alpha substantially improves the model's ability to minimize loss, highlighting alpha's critical role in the model's performance optimization.

Panel C's exploration of the attention dimension (r) further refines the model, maintaining loss around the lowest value achieved in Panel B (0.1122), across different values of r (8, 16, 32, 64). This indicates a plateau in performance improvement concerning r, suggesting that once optimal dropout and alpha values are identified, the attention dimension's influence stabilizes. The experimentation across Panels A to C, each rigorously focusing on one parameter at a time, exemplifies a strategic approach to model optimization. This iterative process not only fine-tunes the model with precision but also achieves a substantial reduction in loss, demonstrating the potential of the proposed fine-tuning strategy outlined in Algorithm 1 for enhancing model efficacy and efficiency in real-world applications.

\begin{table}[ht]
    \centering
    \resizebox{.5\textwidth}{!}{
        \begin{tabular}{cccccc}
        \hline
        \textbf{Panel A} \\
        r & alpha & dropout & epoch & loss & time \\
        \hline
        \multirow{3}{*}{64} & \multirow{3}{*}{16} & 0.001 & \multirow{3}{*}{10} & 0.316 & 6 min 51 sec  \\
         & & 0.01 & & 0.3169 & 6 min 47 sec \\
         & & 0.1 & & 0.3212 & 6 min 49 sec \\
        \hline
        \hline
        \textbf{Panel B} \\
        r & alpha & dropout & epoch & loss & time \\
        \hline
        \multirow{3}{*}{64} & 8 & \multirow{3}{*}{0.001} & \multirow{3}{*}{10} & 0.5904 & 6 min 48 sec  \\
         & 16 & & & 0.316 & 6 min 51 sec \\
         & 32 & & & 0.1585 & 6 min 46 sec \\
         & 64 & & & 0.1122 & 6 min 50 sec \\
        \hline
        \hline
        \textbf{Panel C} \\
        r & alpha & dropout & epoch & loss & time \\
        \hline
        8 & \multirow{3}{*}{64} & \multirow{3}{*}{0.001} & \multirow{3}{*}{10} & 0.1127 & 6 min 45 sec  \\
        16 & & & & 0.1145 & 6 min 41 sec \\
        32 & & & & 0.1122 & 6 min 50 sec \\
        64 & & & & 0.1122 & 7 min 1 sec \\
        \hline
        \end{tabular}
    }
    \caption{\textbf{Fine-tuning results}. This table presents the fine-tuning results according to Algorithm \ref{algo:fine-tune-qlora}. Each panel tunes one parameter and the best parameter is selected when entering into the experiments in the next panel. All unit experiments are ran using 10 epochs and on average it runs a little under 7 minutes. From Panel A to Panel C, we attempt to fine tune each parameter and even with 10 epochs we reduced loss from the original 0.32 to 0.1, a 69\% reduction for iteration of the proposed Algorithm \ref{algo:fine-tune-qlora}. With sufficient computing power, we recommend scholars to repeat the panel many times.}
    \label{tab:fine-tuning-results}
\end{table}

\section{Conclusion}

The research meticulously outlines a novel approach to significantly enhance the capabilities of retrieval-augmented generation (RAG) systems through the integration of fine-tuned large language models (LLMs) with vector databases, leveraging the strengths of both structured data retrieval and advanced LLM understanding. The deployment of LoRA and QLoRA methodologies exemplifies innovative strategies for model refinement, demonstrating the importance of parameter-efficient fine-tuning and memory optimization techniques. This study's inclusion of user feedback into the training loop marks a pivotal advancement, ensuring the model's evolution aligns with user expectations, thereby enhancing its performance and relevance. The introduction of a Quantized Influence Measure (QIM) as an "AI Judge" further sophisticates the model, refining result selection and accuracy. The executive diagram and accompanying algorithm for fine-tuning QLoRA provide a comprehensive blueprint for implementing these advancements within chatbot frameworks, promising significant improvements in chatbot responsiveness and accuracy. This research not only contributes valuable insights into the optimization of LLMs for specific applications but also opens new avenues for further exploration in the enhancement of retrieval-augmented models. Through rigorous experimentation and analysis, the study lays a solid foundation for future advancements in chatbot technology and retrieval systems, signifying a leap forward in the development of more sophisticated, accurate, and user-tailored conversational AI systems.

Building on the promising outcomes of our study, the proposed system offers a broad spectrum of potential applications beyond the immediate focus on enhancing communication within homeless shelters at Youth Spirit Artworks (YSA). This adaptive and robust framework has the potential to revolutionize communication channels across various communities, particularly those in dire need of support, such as low-income groups, educational institutions in underserved areas, and healthcare providers in resource-limited settings. The versatility and efficiency of the system make it a valuable tool for improving access to information, resources, and support, thereby fostering inclusivity and empowerment among vulnerable populations. Furthermore, the technological advancements and methodologies developed through this research have the potential to contribute significantly to the AI community, offering new directions for future innovations in conversational AI and retrieval-augmented systems. We are hopeful that this work will not only pave the way for enhanced communication capabilities within specific communities like YSA but also inspire and facilitate positive impacts on a wider scale, benefiting low-income classes and advancing the field of AI. Through collaborative efforts and continued research, the possibilities for making meaningful, community-driven improvements are limitless, underlining our commitment to leveraging AI for social good and community empowerment.

\section*{Data and Model Availability}

We make the \href{https://huggingface.co/datasets/eagle0504/youthless-homeless-shelter-web-scrape-dataset-large}{dataset} and the \href{https://huggingface.co/eagle0504/llama-2-7b-ysa}{model} publicly available. We also make the app backed the proposed architecture publicly available \href{https://huggingface.co/spaces/eagle0504/YSA-Larkin-Comm}{here}.

\section*{Funding}
No funding information available.

\section*{Authors' contributions}
Keshav Rangan and Yiqiao Yin wrote the main manuscript text. Keshav Rangan and Yiqiao Yin designed the experiment and Keshav Rangan ran the code. Keshav Rangan collected the data and was responsible for the data processing pipeline. Keshav Rangan contributed to the major design of the app backed by the architecture proposed in the paper. Both authors reviewed the manuscript. Both authors read and approved the final manuscript.

\section*{Acknowledgments}
This work is affectionately dedicated to the community of the YSA Homeless Shelter. It is our sincere hope that our ongoing research endeavors will broaden the network of support, extending a helping hand to those in dire need of shelter services.

Our deepest appreciation is extended to Professor Herman Chernoff and Professor Shaw-hwa Lo. Their pioneering work in developing the I-score's theoretical framework and their leadership in preceding research have laid the foundational stones for this critical statistical concept, guiding our path and inspiring our efforts.

\section*{Ethical Declarations}
The authors declare no competing interests.

\bibliographystyle{unsrt}  
\bibliography{references}

\begin{thebibliography}{10}

\bibitem{thompson2002short}
Sanna~J Thompson, David~E Pollio, Jodi Constantine, Donna Reid, and Von Nebbitt.
\newblock Short-term outcomes for youth receiving runaway and homeless shelter services.
\newblock {\em Research on Social Work Practice}, 12(5):589--603, 2002.

\bibitem{spiegler2022crisis}
Jonathan Spiegler, Carlos G{\"u}ereca, Dominic McQuerry, and Erin Troedson.
\newblock From crisis to housing: A comparison of select homeless shelters from across the united states.
\newblock {\em Journal of Poverty}, pages 1--18, 2022.

\bibitem{barber2005homeless}
Carol~Cornsweet Barber, Peter Fonagy, Jim Fultz, Mary~Ann Simulinas, and Miranda Yates.
\newblock Homeless near a thousand homes: Outcomes of homeless youth in a crisis shelter.
\newblock {\em American Journal of Orthopsychiatry}, 75(3):347--355, 2005.

\bibitem{dalton2002adjustment}
Melanie~M Dalton and Kenneth~I Pakenham.
\newblock Adjustment of homeless adolescents to a crisis shelter: Application of a stress and coping model.
\newblock {\em Journal of Youth and Adolescence}, 31:79--89, 2002.

\bibitem{burt2001helping}
Martha~R Burt.
\newblock {\em Helping America's homeless: Emergency shelter or affordable housing?}
\newblock The Urban Insitute, 2001.

\bibitem{dreyer2018shelter}
Benard~P Dreyer.
\newblock A shelter is not a home: The crisis of family homelessness in the united states.
\newblock {\em Pediatrics}, 142(5), 2018.

\bibitem{wallace2018sheltering}
Bruce Wallace, Katrina Barber, and Bernadette~Bernie Pauly.
\newblock Sheltering risks: Implementation of harm reduction in homeless shelters during an overdose emergency.
\newblock {\em International Journal of Drug Policy}, 53:83--89, 2018.

\bibitem{hurtubise2009shelters}
Roch Hurtubise, Pierre-Olivier Babin, and Carolyne Grimard.
\newblock Shelters for the homeless: Learning from research.
\newblock {\em Hulchanski, JD, Campsie, P., Chau, Hwang, S. and Paradis, E.(eds), Finding Home: Policy Options for Addressing Homelessness in Canada, revised edn. Toronto: Cities Centre, University of Toronto}, pages 1--24, 2009.

\bibitem{santos2020elderly}
Fernanda Santos.
\newblock Elderly and homeless: America's next housing crisis.
\newblock {\em New York Times Magazine. Available at: https://www. nytimes. com/2020/09/30/magazine/homeless-seniors-elderly. html}, 2020.

\bibitem{wusinich2019if}
Christina Wusinich, Lynden Bond, Anna Nathanson, and Deborah~K Padgett.
\newblock “if you’re gonna help me, help me”: Barriers to housing among unsheltered homeless adults.
\newblock {\em Evaluation and Program Planning}, 76:101673, 2019.

\bibitem{hocking2000changing}
John~E Hocking and Samuel~G Lawrence.
\newblock Changing attitudes toward the homeless: The effects of prosocial communication with the homeless.
\newblock {\em Journal of social distress and the homeless}, 9:91--110, 2000.

\bibitem{brown2017waiting}
Molly Brown, Martina Mihelicova, Jennifer Lyons, Jennifer DeFonzo, Samantha Torello, Andr{\'e}s Carri{\'o}n, and Allison~N Ponce.
\newblock Waiting for shelter: Perspectives on a homeless shelter's procedures.
\newblock {\em Journal of Community Psychology}, 45(7):846--858, 2017.

\bibitem{greysen2012understanding}
S~Ryan Greysen, Rebecca Allen, Georgina~I Lucas, Emily~A Wang, and Marjorie~S Rosenthal.
\newblock Understanding transitions in care from hospital to homeless shelter: a mixed-methods, community-based participatory approach.
\newblock {\em Journal of general internal medicine}, 27:1484--1491, 2012.

\bibitem{vellozzi2021disparities}
Cristina Vellozzi-Averhoff, William~W Thompson, Claudia Vellozzi, Ike Okosun, Kathy Kinlaw, and Jada Bussey-Jones.
\newblock Disparities in communication among the inpatient homeless population at a safety-net hospital.
\newblock {\em Journal of the National Medical Association}, 113(4):440--448, 2021.

\bibitem{barker1990home}
Robert~L Barker.
\newblock At home with the homeless: An experience in transcultural communication.
\newblock {\em Journal of Independent Social Work}, 4(4):61--73, 1990.

\bibitem{haag2011impacting}
Marcy Haag, Tom Wood, and Linda Holloway.
\newblock Impacting quality of life at a homeless shelter: Measuring the effectiveness of say it straight.
\newblock {\em International Journal of Interdisciplinary Social Sciences}, 5(12), 2011.

\bibitem{olufemi2002barriers}
Olusola Olufemi.
\newblock Barriers that disconnect homeless people and make homelessness difficult to interpret.
\newblock {\em Development Southern Africa}, 19(4):455--466, 2002.

\bibitem{haupt2023examining}
Brittany~“Brie” Haupt and Karen~D Sweeting.
\newblock Examining communication for homeless populations in times of crises.
\newblock {\em Natural Hazards Review}, 24(3):05023003, 2023.

\bibitem{he2023large}
Zhankui He, Zhouhang Xie, Rahul Jha, Harald Steck, Dawen Liang, Yesu Feng, Bodhisattwa~Prasad Majumder, Nathan Kallus, and Julian McAuley.
\newblock Large language models as zero-shot conversational recommenders.
\newblock In {\em Proceedings of the 32nd ACM international conference on information and knowledge management}, pages 720--730, 2023.

\bibitem{brown2020language}
Tom Brown, Benjamin Mann, Nick Ryder, Melanie Subbiah, Jared~D Kaplan, Prafulla Dhariwal, Arvind Neelakantan, Pranav Shyam, Girish Sastry, Amanda Askell, et~al.
\newblock Language models are few-shot learners.
\newblock {\em Advances in neural information processing systems}, 33:1877--1901, 2020.

\bibitem{babaei2023llms4ol}
Hamed Babaei~Giglou, Jennifer D’Souza, and S{\"o}ren Auer.
\newblock Llms4ol: Large language models for ontology learning.
\newblock In {\em International Semantic Web Conference}, pages 408--427. Springer, 2023.

\bibitem{winograd2023loose}
Amy Winograd.
\newblock Loose-lipped large language models spill your secrets: The privacy implications of large language models.
\newblock {\em Harvard Journal of Law \& Technology}, 36(2), 2023.

\bibitem{yang2023fingpt}
Hongyang Yang, Xiao-Yang Liu, and Christina~Dan Wang.
\newblock Fingpt: Open-source financial large language models.
\newblock {\em arXiv preprint arXiv:2306.06031}, 2023.

\bibitem{ferber2024large}
Dyke Ferber and Jakob~Nikolas Kather.
\newblock Large language models in uro-oncology.
\newblock {\em European Urology Oncology}, 7(1):157--159, 2024.

\bibitem{ozdemir2023quick}
Sinan Ozdemir.
\newblock {\em Quick Start Guide to Large Language Models: Strategies and Best Practices for Using ChatGPT and Other LLMs}.
\newblock Addison-Wesley Professional, 2023.

\bibitem{jamal2023improved}
Suhaima Jamal and Hayden Wimmer.
\newblock An improved transformer-based model for detecting phishing, spam, and ham: A large language model approach.
\newblock {\em arXiv preprint arXiv:2311.04913}, 2023.

\bibitem{pan2023integrating}
Shirui Pan, Yizhen Zheng, and Yixin Liu.
\newblock Integrating graphs with large language models: Methods and prospects.
\newblock {\em arXiv preprint arXiv:2310.05499}, 2023.

\bibitem{kumar2023large}
Vimal Kumar, Priyam Srivastava, Ashay Dwivedi, Ishan Budhiraja, Debjani Ghosh, Vikas Goyal, and Ruchika Arora.
\newblock Large-language-models (llm)-based ai chatbots: Architecture, in-depth analysis and their performance evaluation.
\newblock In {\em International Conference on Recent Trends in Image Processing and Pattern Recognition}, pages 237--249. Springer, 2023.

\bibitem{rasnayaka2024empirical}
Sanka Rasnayaka, Guanlin Wang, Ridwan Shariffdeen, and Ganesh~Neelakanta Iyer.
\newblock An empirical study on usage and perceptions of llms in a software engineering project.
\newblock {\em arXiv preprint arXiv:2401.16186}, 2024.

\bibitem{levy2023guiding}
Mosh Levy, Shauli Ravfogel, and Yoav Goldberg.
\newblock Guiding llm to fool itself: Automatically manipulating machine reading comprehension shortcut triggers.
\newblock {\em arXiv preprint arXiv:2310.18360}, 2023.

\bibitem{deng2023efficient}
Zhijie Deng, Hongcheng Gao, Yibo Miao, and Hao Zhang.
\newblock Efficient detection of llm-generated texts with a bayesian surrogate model.
\newblock {\em arXiv preprint arXiv:2305.16617}, 2023.

\bibitem{ge2023openagi}
Yingqiang Ge, Wenyue Hua, Jianchao Ji, Juntao Tan, Shuyuan Xu, and Yongfeng Zhang.
\newblock Openagi: When llm meets domain experts.
\newblock {\em arXiv preprint arXiv:2304.04370}, 2023.

\bibitem{xue2023repeat}
Fuzhao Xue, Yao Fu, Wangchunshu Zhou, Zangwei Zheng, and Yang You.
\newblock To repeat or not to repeat: Insights from scaling llm under token-crisis.
\newblock {\em arXiv preprint arXiv:2305.13230}, 2023.

\bibitem{bekbayev2023poison}
Aibek Bekbayev, Sungbae Chun, Yerzat Dulat, and James Yamazaki.
\newblock The poison of alignment.
\newblock {\em arXiv preprint arXiv:2308.13449}, 2023.

\bibitem{dettmers2023qlora}
Tim Dettmers, Artidoro Pagnoni, Ari Holtzman, and Luke Zettlemoyer.
\newblock Qlora: Efficient finetuning of quantized llms.
\newblock {\em arXiv preprint arXiv:2305.14314}, 2023.

\bibitem{li2023loftq}
Yixiao Li, Yifan Yu, Chen Liang, Pengcheng He, Nikos Karampatziakis, Weizhu Chen, and Tuo Zhao.
\newblock Loftq: Lora-fine-tuning-aware quantization for large language models.
\newblock {\em arXiv preprint arXiv:2310.08659}, 2023.

\bibitem{zhang2024quantized}
Zhengxin Zhang, Dan Zhao, Xupeng Miao, Gabriele Oliaro, Qing Li, Yong Jiang, and Zhihao Jia.
\newblock Quantized side tuning: Fast and memory-efficient tuning of quantized large language models.
\newblock {\em arXiv preprint arXiv:2401.07159}, 2024.

\bibitem{jeon2024l4q}
Hyesung Jeon, Yulhwa Kim, and Jae-joon Kim.
\newblock L4q: Parameter efficient quantization-aware training on large language models via lora-wise lsq.
\newblock {\em arXiv preprint arXiv:2402.04902}, 2024.

\bibitem{yin2023modulora}
Junjie Yin, Jiahao Dong, Yingheng Wang, Christopher De~Sa, and Volodymyr Kuleshov.
\newblock Modulora: Finetuning 3-bit llms on consumer gpus by integrating with modular quantizers.
\newblock {\em arXiv preprint arXiv:2309.16119}, 2023.

\bibitem{zhang2023machine}
Xuan Zhang, Navid Rajabi, Kevin Duh, and Philipp Koehn.
\newblock Machine translation with large language models: Prompting, few-shot learning, and fine-tuning with qlora.
\newblock In {\em Proceedings of the Eighth Conference on Machine Translation}, pages 468--481, 2023.

\bibitem{xu2023qa}
Yuhui Xu, Lingxi Xie, Xiaotao Gu, Xin Chen, Heng Chang, Hengheng Zhang, Zhensu Chen, Xiaopeng Zhang, and Qi~Tian.
\newblock Qa-lora: Quantization-aware low-rank adaptation of large language models.
\newblock {\em arXiv preprint arXiv:2309.14717}, 2023.

\bibitem{guo2023lq}
Han Guo, Philip Greengard, Eric~P Xing, and Yoon Kim.
\newblock Lq-lora: Low-rank plus quantized matrix decomposition for efficient language model finetuning.
\newblock {\em arXiv preprint arXiv:2311.12023}, 2023.

\bibitem{weng2023lmtuner}
Yixuan Weng, Zhiqi Wang, Huanxuan Liao, Shizhu He, Shengping Liu, Kang Liu, and Jun Zhao.
\newblock Lmtuner: An user-friendly and highly-integrable training framework for fine-tuning large language models.
\newblock {\em arXiv preprint arXiv:2308.10252}, 2023.

\bibitem{lewis2020retrieval}
Patrick Lewis, Ethan Perez, Aleksandra Piktus, Fabio Petroni, Vladimir Karpukhin, Naman Goyal, Heinrich K{\"u}ttler, Mike Lewis, Wen-tau Yih, Tim Rockt{\"a}schel, et~al.
\newblock Retrieval-augmented generation for knowledge-intensive nlp tasks.
\newblock {\em Advances in Neural Information Processing Systems}, 33:9459--9474, 2020.

\bibitem{mao2020generation}
Yuning Mao, Pengcheng He, Xiaodong Liu, Yelong Shen, Jianfeng Gao, Jiawei Han, and Weizhu Chen.
\newblock Generation-augmented retrieval for open-domain question answering.
\newblock {\em arXiv preprint arXiv:2009.08553}, 2020.

\bibitem{cai2022recent}
Deng Cai, Yan Wang, Lemao Liu, and Shuming Shi.
\newblock Recent advances in retrieval-augmented text generation.
\newblock In {\em Proceedings of the 45th International ACM SIGIR Conference on Research and Development in Information Retrieval}, pages 3417--3419, 2022.

\bibitem{liu2020retrieval}
Shangqing Liu, Yu~Chen, Xiaofei Xie, Jingkai Siow, and Yang Liu.
\newblock Retrieval-augmented generation for code summarization via hybrid gnn.
\newblock {\em arXiv preprint arXiv:2006.05405}, 2020.

\bibitem{gao2023retrieval}
Yunfan Gao, Yun Xiong, Xinyu Gao, Kangxiang Jia, Jinliu Pan, Yuxi Bi, Yi~Dai, Jiawei Sun, and Haofen Wang.
\newblock Retrieval-augmented generation for large language models: A survey.
\newblock {\em arXiv preprint arXiv:2312.10997}, 2023.

\bibitem{jiang2023active}
Zhengbao Jiang, Frank~F Xu, Luyu Gao, Zhiqing Sun, Qian Liu, Jane Dwivedi-Yu, Yiming Yang, Jamie Callan, and Graham Neubig.
\newblock Active retrieval augmented generation.
\newblock {\em arXiv preprint arXiv:2305.06983}, 2023.

\bibitem{kim2020retrieval}
Jihyeok Kim, Seungtaek Choi, Reinald~Kim Amplayo, and Seung-won Hwang.
\newblock Retrieval-augmented controllable review generation.
\newblock In {\em Proceedings of the 28th International Conference on Computational Linguistics}, pages 2284--2295, 2020.

\bibitem{chen2023benchmarking}
Jiawei Chen, Hongyu Lin, Xianpei Han, and Le~Sun.
\newblock Benchmarking large language models in retrieval-augmented generation.
\newblock {\em arXiv preprint arXiv:2309.01431}, 2023.

\bibitem{li2022survey}
Huayang Li, Yixuan Su, Deng Cai, Yan Wang, and Lemao Liu.
\newblock A survey on retrieval-augmented text generation.
\newblock {\em arXiv preprint arXiv:2202.01110}, 2022.

\bibitem{goyal2022retrieval}
Anirudh Goyal, Abram Friesen, Andrea Banino, Theophane Weber, Nan~Rosemary Ke, Adria~Puigdomenech Badia, Arthur Guez, Mehdi Mirza, Peter~C Humphreys, Ksenia Konyushova, et~al.
\newblock Retrieval-augmented reinforcement learning.
\newblock In {\em International Conference on Machine Learning}, pages 7740--7765. PMLR, 2022.

\bibitem{blattmann2022retrieval}
Andreas Blattmann, Robin Rombach, Kaan Oktay, Jonas M{\"u}ller, and Bj{\"o}rn Ommer.
\newblock Retrieval-augmented diffusion models.
\newblock {\em Advances in Neural Information Processing Systems}, 35:15309--15324, 2022.

\bibitem{siriwardhana2023improving}
Shamane Siriwardhana, Rivindu Weerasekera, Elliott Wen, Tharindu Kaluarachchi, Rajib Rana, and Suranga Nanayakkara.
\newblock Improving the domain adaptation of retrieval augmented generation (rag) models for open domain question answering.
\newblock {\em Transactions of the Association for Computational Linguistics}, 11:1--17, 2023.

\bibitem{gao2022retrieval}
Yifan Gao, Qingyu Yin, Zheng Li, Rui Meng, Tong Zhao, Bing Yin, Irwin King, and Michael~R Lyu.
\newblock Retrieval-augmented multilingual keyphrase generation with retriever-generator iterative training.
\newblock {\em arXiv preprint arXiv:2205.10471}, 2022.

\bibitem{guo2024retrieval}
Yue Guo, Wei Qiu, Gondy Leroy, Sheng Wang, and Trevor Cohen.
\newblock Retrieval augmentation of large language models for lay language generation.
\newblock {\em Journal of Biomedical Informatics}, 149:104580, 2024.

\bibitem{chernoffetal2009}
Herman Chernoff, Shaw-Hwa Lo, and Tian Zheng.
\newblock {Discovering influential variables: A method of partitions}.
\newblock {\em The Annals of Applied Statistics}, 3(4):1335 -- 1369, 2009.

\bibitem{lozheng2002}
S.H. Lo and T.~Zheng.
\newblock Backward haplotype transmission association algorithm - a fast multiple-marker screening method.
\newblock {\em Hum. Hered.}, 53(4):197--215, 2002.

\bibitem{lo2021interaction}
Shaw-Hwa Lo and Yiqiao Yin.
\newblock An interaction-based convolutional neural network (icnn) toward a better understanding of covid-19 x-ray images.
\newblock {\em Algorithms}, 14(11):337, 2021.

\bibitem{lo2021novel}
Shaw-Hwa Lo and Yiqiao Yin.
\newblock A novel interaction-based methodology towards explainable ai with better understanding of pneumonia chest x-ray images.
\newblock {\em Discover Artificial Intelligence}, 1(1):16, 2021.

\bibitem{lo2021language}
Shaw-Hwa Lo and Yiqiao Yin.
\newblock Language semantics interpretation with an interaction-based recurrent neural network.
\newblock {\em Machine Learning and Knowledge Extraction}, 3(4):922--945, 2021.

\bibitem{di2023detecting}
Xuan Di, Yiqiao Yin, Yongjie Fu, Zhaobin Mo, Shaw-Hwa Lo, Carolyn DiGuiseppi, David~W Eby, Linda Hill, Thelma~J Mielenz, David Strogatz, et~al.
\newblock Detecting mild cognitive impairment and dementia in older adults using naturalistic driving data and interaction-based classification from influence score.
\newblock {\em Artificial Intelligence in Medicine}, 138:102510, 2023.

\bibitem{lo2015significant}
Adeline Lo, Herman Chernoff, Tian Zheng, and Shaw-Hwa Lo.
\newblock Why significant variables aren’t automatically good predictors.
\newblock {\em Proceedings of the National Academy of Sciences}, 112(45):13892--13897, 2015.

\bibitem{lo2016framework}
Adeline Lo, Herman Chernoff, Tian Zheng, and Shaw-Hwa Lo.
\newblock Framework for making better predictions by directly estimating variables’ predictivity.
\newblock {\em Proceedings of the National Academy of Sciences}, 113(50):14277--14282, 2016.

\bibitem{aghajanyan2020intrinsic}
Armen Aghajanyan, Luke Zettlemoyer, and Sonal Gupta.
\newblock Intrinsic dimensionality explains the effectiveness of language model fine-tuning.
\newblock {\em arXiv preprint arXiv:2012.13255}, 2020.

\bibitem{he2023efficientdm}
Yefei He, Jing Liu, Weijia Wu, Hong Zhou, and Bohan Zhuang.
\newblock Efficientdm: Efficient quantization-aware fine-tuning of low-bit diffusion models.
\newblock {\em arXiv preprint arXiv:2310.03270}, 2023.

\bibitem{schreiber2023esmbind}
Amelie Schreiber.
\newblock Esmbind and qbind: Lora, qlora, and esm-2 for predicting binding sites and post translational modification.
\newblock {\em bioRxiv}, pages 2023--11, 2023.

\bibitem{zi2023delta}
Bojia Zi, Xianbiao Qi, Lingzhi Wang, Jianan Wang, Kam-Fai Wong, and Lei Zhang.
\newblock Delta-lora: Fine-tuning high-rank parameters with the delta of low-rank matrices.
\newblock {\em arXiv preprint arXiv:2309.02411}, 2023.

\bibitem{xia2024chain}
Wenhan Xia, Chengwei Qin, and Elad Hazan.
\newblock Chain of lora: Efficient fine-tuning of language models via residual learning.
\newblock {\em arXiv preprint arXiv:2401.04151}, 2024.

\end{thebibliography}

\end{document}